\documentclass[11pt,a4paper]{article}

\usepackage{geometry}
\usepackage{amsmath}
\usepackage{amsfonts}
\usepackage[T1]{fontenc}
\usepackage{graphicx}

\newcommand{\beq}{\begin{equation}}
\newcommand{\eeq}{\end{equation}}
\newcommand{\bal}{\begin{aligned}}
\newcommand{\eal}{\end{aligned}}
\newcommand{\SL}{SL(2,\mathbb{R})}
\newcommand{\scri}{\mathcal{J}}

\begin{document}

\title{How trapped surfaces jump in 2+1 dimensions}
\author{Emma Jakobsson\footnote{emma.jakobsson@fysik.su.se}}
\date{\normalsize{\emph{Fysikum, Stockholms Universitet,}}\\
	\normalsize{\emph{S-106 91, Stockholm, Sweden}}}
\maketitle

\begin{abstract}

When a lump of matter falls into a black hole it is expected that a marginally trapped tube when hit moves outwards everywhere, even in regions not yet in causal contact with the infalling matter. But to describe this phenomenon analytically in 3+1 dimensions is difficult since gravitational radiation is emitted. By considering a particle falling into a toy model of a black hole in 2+1 dimensions an exact description of this non-local behaviour of a marginally trapped tube is found.

\end{abstract}

\section{Introduction}
A black hole is defined by its event horizon; a boundary in spacetime, such that no event inside it can ever be seen from the outside. With this definition it is impossible to locate the event horizon without knowledge about the infinite future. Attempts to make alternative definitions of a black hole involve trapped surfaces that occur in the interior \cite{hayward02,ashtekar04,booth05}. A trapped surface is a closed, spacelike surface such that both families of light rays orthogonal to it converge. The terminology of concepts closely related to these trapped surfaces might need to be made clear: A closed spacelike surface such that only one of the orthogonal families of light rays converges while the other has zero convergence, is referred to as a \emph{marginally} trapped surface. If the surface is embedded in a hypersurface on which an outer direction is defined in a manner that would be intuitive in an asymptotically simple spacetime, and this surface is such that the outgoing family of light rays orthogonal to it converges, it is called \emph{outer} trapped, regardless of the behaviour of the ingoing family of light rays. \emph{Marginally outer} trapped surfaces are defined in a similar manner. While the event horizon is a globally defined property of spacetime -- and therefore, as we will see, teleological in its nature -- trapped surfaces are quasilocal, since their definition only involves the surfaces themselves and their infinitesimal surroundings. For this reason trapped surfaces are of importance to numerical relativists, since the occurence of such is the only practical way to identify a black hole in a simulated evolution of spacelike hypersurfaces. In such simulations the trapped surfaces sometimes make discontinuous ``jumps'' outwards \cite{andersson05,jaramillo09}. This phenomenon is expected when matter is falling into the black hole \cite{hawking73}. 

A \emph{marginally trapped tube} is a hypersurface foliated by marginally trapped surfaces. The marginally trapped tubes we will come across will be null and satisfy some other constraints that qualify them as isolated horizons \cite{ashtekar02}. It is desirable to find an exact description of how a marginally trapped tube is affected when hit by matter. This problem has also been studied in spherically symmetric cases \cite{ben-dov04,booth06}. However, if a localized ``lump'' of matter is falling into a black hole, it is much more difficult to find an analytical description since gravitational radiation is emitted. But it is expected that the jump in this case will be in some sense non-local; that the jump will take place also in regions not yet in causal contact with the infalling matter. There is no need to worry about causality violation; this effect is just a consequence of the quasilocal definition of a trapped surface. Light rays emitted from a region on a spacelike surface may converge, but whether the whole surface is closed -- and thus trapped -- or not depends on circumstances elsewhere.

Because of the difficulties in 3+1 dimensions we instead tackle the problem in 2+1 dimensions \cite{deserjackiw84} where there is no gravitational radiation. We consider a toy model of a black hole and let a point particle fall into it in order to find an exact description of how the marginally trapped tube jumps outwards in this non-local way.

\section{The black hole and trapped surfaces}
The existence of a black hole in a 2+1-dimensional spacetime with constant negative curvature was first discovered by Ba\~{n}ados \emph{et al} \cite{banados92}. This is called a BTZ black hole. It is obtained by identifying points in anti-de Sitter space using an isometry \cite{banados93}.

2+1-dimensional anti-de Sitter space can be defined as the hypersurface
\beq
	X^2+Y^2-U^2-V^2=-1,
\eeq
embedded in a four dimensional spacetime with metric
\beq
	ds^2=dX^2+dY^2-dU^2-dV^2.
\eeq
It has constant curvature which is negative. Each point can be represented by a matrix
\beq
	\boldsymbol{g}=\begin{pmatrix}
		U+Y & X+V\\
		X-V & U-Y
		\end{pmatrix},
\eeq
so that
\beq
	\det\boldsymbol{g}=-X^2-Y^2+U^2+V^2=1.
\eeq
But this is a group element of $\SL$, consisting of all two by two matrices with real matrix elements and determinant one. Furthermore, any isometry can be described by letting the group act on itself. Isometries leaving the unit element fixed can be written
\beq \label{transformation}
	\boldsymbol{g}\rightarrow\boldsymbol{g}'=\boldsymbol{g}_1\boldsymbol{g}\boldsymbol{g}_1^{-1},
\eeq
where $\boldsymbol{g}_1\in\SL$. Transformations of the type \eqref{transformation} will have a line of fixed points and the nature of this line is determined by the trace of $\boldsymbol{g}_1$. If Tr$\boldsymbol{g}_1<2$ it will be timelike, if Tr$\boldsymbol{g}_1=2$ it will be lightlike and if Tr$\boldsymbol{g}_1>2$ it will be spacelike.

The embedding coordinates are convenient to use in calculations, but for visualization the intrinsic coordinates $(t,\rho,\phi)$ \cite{holst00} are a better choice. They are given by
\beq \label{sausage}
	\bal
		X & =\frac{2\rho}{1-\rho^2}\cos\phi\\
		Y & =\frac{2\rho}{1-\rho^2}\sin\phi\\
		U & =\frac{1+\rho^2}{1-\rho^2}\cos t\\
		V & =\frac{1+\rho^2}{1-\rho^2}\sin t
	\eal
	\qquad \qquad
	\bal
		& 0\leq\rho<1 \\
		& 0\leq\phi<2\pi \\
		& -\pi\leq t<\pi.
	\eal
\eeq
The metric in these coordinates is
\beq \label{ds}
	ds^2=-\left(\frac{1+\rho^2}{1-\rho^2}\right)^2dt^2+\frac{4}{(1-\rho^2)^2}(d\rho^2+\rho^2d\phi^2).
\eeq
With this choice of coordinates anti-de Sitter space is depicted as a cylinder. The timelike coordinate $t$ runs along the cylinder, and the spatial slices of constant $t$ are Poincar\'{e} disks. On the disk, $\rho$ and $\phi$ are the radial and angular coordinates respectively and $\scri$ is situated at the boundary $\rho=1$.

\begin{figure}[ht]
	\begin{center}
	\leavevmode
	\includegraphics[width=80mm]{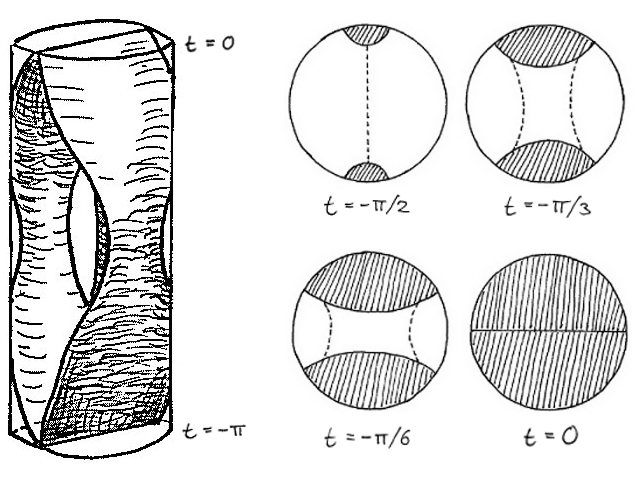}
	\end{center}
	\caption{\small The BTZ black hole. The cylinder is depicting 2+1-dimensional anti-de Sitter space in which the identification surfaces are drawn. To the right are spatial slices with different values of constant $t$. As the identification is performed the shaded regions are cut away, and each slice, except $t=0$, turns into a cylinder with two asymptotic regions. In this figure, only the top left disk where $t=-\pi/2$ will turn into a smooth surface by the identification, since the flow lines of the identification in general do not lie on a disk of constant $t$. But the full spacetime is smooth everywhere except at the singularity, drawn on the bottom right disk where $t=0$. The dashed curves on the disks are the event horizons -- one for each asymptotic region. (This figure is a paraphrase on a figure originally drawn by S\"{o}ren Holst \cite{aminneborg96}.)}
	\label{soren}
\end{figure}

To create a black hole we choose a group element
\beq \label{gbh}
	\boldsymbol{g}_{BH}=\begin{pmatrix}
	\cosh\mu & \sinh\mu\\
	\sinh\mu & \cosh\mu
	\end{pmatrix}.
\eeq
The real constant $\mu$ will determine the mass of the black hole. Then we act with $\boldsymbol{g}_{BH}$ on anti-de Sitter space through conjugation as in Eq. \eqref{transformation}, and identify points that are transformed into each other. The region between the two surfaces $Y=V\tanh\mu$ and $Y=-V\tanh\mu$ can be taken to represent the resulting quotient space, as in Fig. \ref{soren}. Due to the identification a spacelike slice now has the geometry of a cylinder, but space is still locally anti-de Sitter everywhere. Note that there are two asymptotic regions, as in the Schwarzschild solution in which one of the regions is considered unphysical. The fixed points of the transformation yielding the identification are located at the spacelike line $Y=V=0$. Starting from the slice $t=-\pi/2$ it is seen that the cylinders shrink in the periodical direction as $t$ increases, until one dimension suddenly disappears at $t=0$, and all that is left is the line of fixed points. A geodesic ending at this singular line ends after only a finite parameter time, meaning that this spacetime is geodesically incomplete. The event horizon is the backward light cone of the last point on $\scri$, i.e. the point where the singular line meets $\scri$. There is one event horizon for each asymptotic region. In the embedding coordinates the event horizons are given as the quotient of each of the two surfaces $X=\pm U$.

The black hole spacetime is locally anti-de Sitter everywhere except at the singular line. On a spacelike surface, the only way to distinguish it from anti-de Sitter space is through the holonomy of the black hole: If a vector is parallel transported along a curve closed by the identification it will also be transformed by the group element effecting the identification.

Finding trapped surfaces -- or rather trapped curves, since we are in 2+1 dimensions -- is easy. Consider the intersection of two light cones with vertices at the singularity. Light rays emanating orthogonally from such curves obviously converge. Moreover they coincide with flow lines of the identifying isometry and are therefore closed to smooth curves by the identification. Hence they are trapped. By letting one of the two vertices be on $\scri$, and varying the other, it is easily seen that the event horizon is a marginally trapped tube, that is a surface foliated by marginally trapped curves. Since trapped surfaces can not exist outside the event horizon according to the cosmic censorship hypothesis, the marginally trapped tube -- that is the event horizon in this model -- is also the boundary of the region containing trapped curves.

In fact this is the complete picture: all marginally trapped curves lie on the event horizon. To see this, consider Raychaudhuri's equation \cite{wald84} for the expansion $\theta$ of a congruence of lightlike geodesics in 2+1 dimensions. With $k^a$ being the tangent vector of a given geodesic we have
\beq \label{raychaudhuri}
	\dot{\theta} = -\theta^2-R_{ab}k^ak^b.
\eeq
If we impose Einstein's vacuum equation $R_{ab}=\lambda g_{ab}$ the second term vanishes since $k^2=0$ for a lightlike geodesic. We are left with
\beq
	\dot{\theta} = -\theta^2,
\eeq
which shows that a congruence of lightlike geodesics that have zero convergence at some point, must continue to have zero convergence. The conclusion is that a marginally trapped curve must lie on a null plane\footnote{This does not hold in 3+1 dimensions, since Raychaudhuri's equation will then contain extra terms.}, where a null plane is defined as a light cone with its vertex on $\scri$. It is not difficult to show that only the null plane containing a fixed point on $\scri$ contains smooth and spacelike closed curves.

As a side note, there is a theorem that says that a region of a spacelike hypersurface bounded by an outer trapped surface in one direction and by an outer untrapped surface in the other must contain a marginally outer trapped surface \cite{andersson09}. In this model the statement is almost obvious. Any smooth spacelike surface passing through the interior of the black hole will contain a smooth closed curve lying on the event horizon and thus being a marginally outer trapped curve. Since it lies on the event horizon it also separates the region containing trapped curves from the region not containing trapped curves on the surface.

\section{The infalling particle}
Just like a black hole was obtained by identifying points, a point particle can be modelled using the same trick. Note that the matrix of Eq. \eqref{gbh} has a trace larger than two, and therefore has a spacelike line of fixed points. If we instead choose the group element
\beq \label{gp}
	\boldsymbol{g}_P=\begin{pmatrix}
		1 & 2a\\
		0 & 1
	\end{pmatrix},
\eeq
with $a$ being an arbitrary real constant, and identify points in anti-de Sitter space through conjugation, the line of fixed points will be lightlike since Tr$\boldsymbol{g}_P=2$. A fundamental region containing one representative of every point in the quotient space can be chosen by cutting away the wedge between the two identified surfaces $Y=\pm a(X-V)$. The effect is that a surface of constant $t$ now has the geometry of a cone, with the tip of the cone being a fixed point of the identification. This setup perfectly well describes a point particle \cite{deser84,deser92}. The particle is situated at the conical singularity, and it is a lightlike particle since its world line is lightlike. Let us consider a sequence of Poincar\'{e} disks. Before the time $t=-\pi/2$ there is no particle, just empty anti-de Sitter space. At $t=-\pi/2$ the particle comes in from infinity. Then it traverses the disk as $t$ increases until it finally leaves at $t=\pi/2$ and we again are left with empty anti-de Sitter space. On the disk, space is locally anti-de Sitter everywhere except at the singularity, and the only way to notice the presence of the particle is to travel around it and reveal its holonomy. That the particle enters empty anti-de Sitter space from infinity is a property unique for lightlike particles in this construction. It is not crucial that the particle we use is lightlike, we might just as well consider a timelike particle. But the advantage of using a lightlike particle is that the starting point will be an undisturbed BTZ spacetime, instead of a white hole emitting massive particles.

\begin{figure}[ht]
	\begin{center}
	\leavevmode
	\includegraphics[width=120mm]{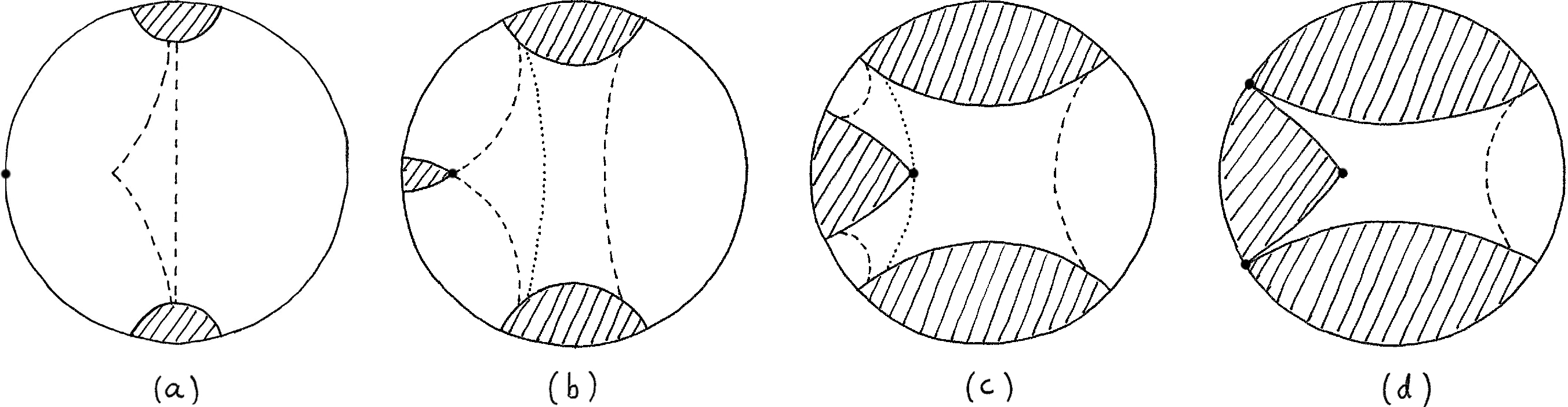}
	\end{center}
	\caption{\small A sequence of Poincar\'{e} disks shows what happens when the particle falls into the black hole (compare with Fig. \ref{fyrkant}). (\emph{a}) The particle comes in from infinity. The event horizon has a kink and does not contain any marginally trapped curves. (\emph{b}) The particle meets the event horizon which from here on is a smooth marginally trapped tube. The dotted curve is the isolated horizon that would have been the event horizon had the particle not been there. (\emph{c}) The isolated horizon in the inner region is hit by the particle. From this point on it ceases to be a smooth marginally trapped tube. (\emph{d}) A fixed point appears on $\scri$ as the identification surfaces of the particle and the black hole begin to intersect. The dashed curve to the right is not relevant in these figures; it is just an artefact of the other asymptotic region.}
	\label{skivor}
\end{figure}

We are now ready to set up a model in which we let the particle fall into the black hole. The result is illustrated in Fig. \ref{skivor}. As the lightlike particle approaches the center of the disk it is seen how the identification surfaces of the particle eventually begin to intersect the identification surfaces of the black hole. These points of intersection are fixed points under the action of the combined holonomy $\boldsymbol{g}_{tot}=\boldsymbol{g}_P\boldsymbol{g}_{BH}$. Here the constants $a$ and $\mu$ are chosen so that |Tr$\boldsymbol{g}_{tot}|>2$ and consequently the transformation $\boldsymbol{g}\rightarrow\boldsymbol{g}_{tot}\boldsymbol{g}\boldsymbol{g}_{tot}^{-1}$ has a spacelike line of fixed points. This spacelike line is singular and appears at smaller $t$ than the singularity of the original black hole. This means that the role of the original singularity is taken over by this new singular line. In turn this affects the location of the event horizon, shown as the dashed curves in Fig. \ref{skivor}. Also the mass of the black hole has been affected by the infalling particle. The change in mass is determined by the constant $a$.

It turns out that the event horizon in this model has a kink before the particle crosses it. This kink nicely illustrates the teleological nature of the event horizon since it has acquired a kink not because of something that has happened to it in the past, but because of something that will happen to it in the future.

Due to the kink the event horizon is not everywhere smooth, with the consequence that it is not completely foliated by marginally trapped curves. The question now is where the marginally trapped curves are in this model. We know that they are found on null planes and that a null plane is smooth only if it contains a fixed point on $\scri$. It is a crucial fact that the light cone on which the path of the particle lies splits the spacetime into two qualitatively different parts.

In the outer region the holonomy is $\boldsymbol{g}_{tot}$. The event horizon is smooth and it contains the point on $\scri$ that is a fixed point under the action of this holonomy. Therefore it is also foliated by marginally trapped curves. Moreover, the event horizon is the boundary of the region containing trapped curves since these can only appear in the interior of the black hole.

In the inner region, on the other hand, the holonomy is $\boldsymbol{g}_{BH}$, and it is therefore isometric to a region of the BTZ spacetime. Restricted to this region, the situation is thus identical to that of a black hole with no infalling particle. All marginally trapped curves lie on the null plane that would have been the event horizon had the particle not been there. And, as we saw, this null plane is also the boundary of the region containing trapped curves. It is an isolated horizon in the terminology of ref. \cite{ashtekar02}, as well as the event horizon in the outer region. But after it has been hit by the particle -- in the outer region -- it is no longer smooth. 

\begin{figure}[ht]
	\begin{center}
	\leavevmode
	\includegraphics[width=100mm]{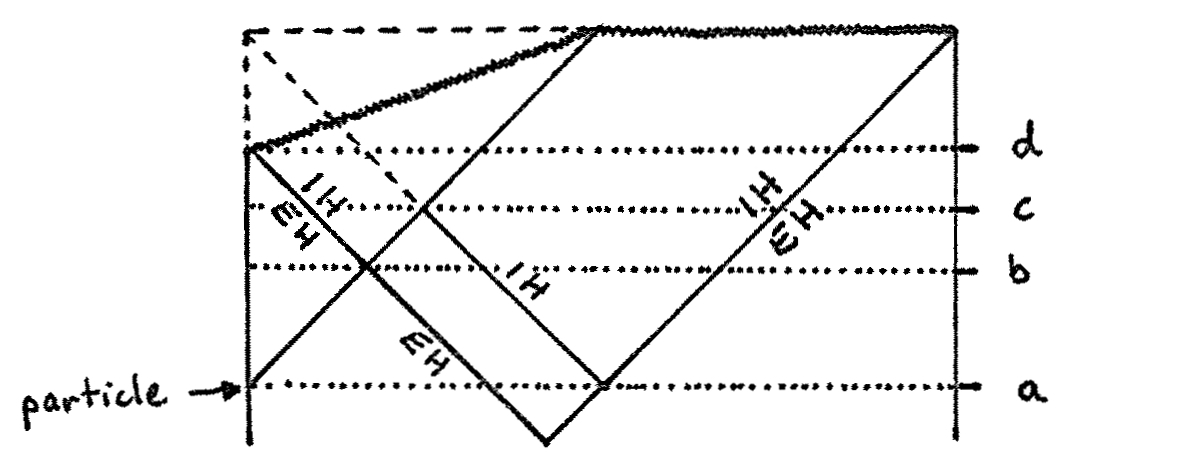}
	\end{center}
	\caption{\small A conformal diagram of our model clearly illustrates how the isolated horizon ``jumps'' outwards when it is hit by the particle. The dashed lines show the location of the singularity and the event horizon had the particle not been there. The light cone on which the path of the particle lies splits the spacetime into two different regions. In the outer region the isolated horizon foliated by marginally trapped curves coincides with the event horizon, and in the inner region it does not.}
	\label{fyrkant}
\end{figure}

The marginally trapped tube thus consists of two parts: the two isolated horizons in the inner and the outer region respectively. All marginally trapped curves lie on the marginally trapped tube, and thus we have a complete knowledge of their whereabouts, independent of a given foliation of spacetime. When the particle hits the isolated horizon in the interior of the black hole, it is seemingly destroyed but then reappears on the event horizon in the outer region, it ``jumps''. This is clearly illustrated in the conformal diagram of Fig. \ref{fyrkant}. With this model in which the marginally trapped tube is discontinuous we have thus found a reasonable and exact illustration of how marginally trapped curves jump when hit by matter.

\section{Conclusions}

By considering a toy model of a black hole in 2+1 dimensions and letting a point particle fall into the black hole, we have seen how the marginally trapped tube splits into two parts. This exact description of the splitting illustrates the non-local jump described in the introduction. Similarly non-local jumps are expected in 3+1 dimensions, but most likely that case must be attacked numerically.

As a concluding remark it is worth noting that since the world line of the particle is singular, the two parts of the marginally trapped tube can not be connected. To get around this problem one could consider a small tube of null dust instead of a point particle. It might be interesting to see what the marginally trapped tube would look like in this more complicated model; in particular if it would be smooth, and if so, if the smooth part joining the two isolated horizons would be timelike or spacelike.

\section*{Acknowledgements}
I would like to thank Ingemar Bengtsson for bringing my attention to the problem and for his significant support, and two anonymous referees for helpful comments. I would also like to thank Sören Holst for accepting the similarities between Fig. \ref{soren} and his original.

\end{document}